# State of the art of Trust and Reputation Systems in E-Commerce Context

H. Rahimi* and H. El Bakkali,
Information Security Research Team, University Mohamed V Rabat ENSIAS, Rabat, Morocco.
h.elbakkali@um5s.net.ma  *hasnae.rahimi@gmail.com

*Abstract*-This article proposes in depth comparative study of the most popular, used and analyzed Trust and Reputation System (TRS) according to the trust and reputation literature and in terms of specific trustworthiness criteria. This survey is realized relying on a selection of trustworthiness criteria that analyze and evaluate the maturity and effectiveness of TRS. These criteria describe the utility, the usability, the performance and the effectiveness of the TRS. We also provide a summary table of the compared TRS within a detailed and granular selection of trust and reputation aspects.

## I. Introduction

Open electronic markets, online collaboration systems, distributed peer-to-peer applications, online social media require the establishment of mutual trust between service providers and service consumers. In fact, the major concerns of web-based services especially e-commerce applications is to overcome the inherent uncertainties and untrustworthiness risks and enhance the system's robustness and resistance against fraudulent users and distrustful ones. Besides, e-commerce platforms aim at adopting the most efficient approach that helps detect and analyze users' intentions in order to reveal and understand deceitful ones. Otherwise, the underlying purpose of e-commerce services which is to maximize the profit and the rate of purchase, would be threatened and deteriorated by fraudulent and ill-intentioned users.

For this reason, Recommender Systems such as Trust and Reputation Systems (TRS), provide essential input for computational trust so as to predict future behaviors of peers basing on the past actions of a peer [1]. In a reputation network, information about these actions can also be received from other members of a reputation network who have transacted with the peer. However, the credibility of this third-party information must be critically assessed. The underlying goal in all reputation systems is to predict a peer's future transactions taking into account his past actions and applying algorithms relying on probabilities approaches.

To gather these first-hand transactions is a tedious and costly task especially when involving the malicious and fraudulent peers. To overcome this limitation, users share their experiences through the reputation system, which aims to detect and effectively isolate misbehaving customers and users. Indeed, e-commerce users refer to this shared information as recommendations on which they rely in order to make the right purchase decision. As no user in e-commerce environment is fully trusted, recommendations and reviews credibility and trustworthiness must be critically assessed. At this purpose, Trust and Reputation Systems have been widely used for various e-commerce applications in order to assess the credibility and trustworthiness of the provided reputation information. Furthermore, this task is becoming increasingly important for the majority of e-services, but especially for e-commerce platforms where resources and business benefice value depend on making correct decisions. As a matter of fact, deliberately providing fake and dishonest ratings and reviews is a serious and crucial security issue that threaten the trust establishment and propagation in e-commerce environment. In fact, this misbehaving attitude would systematically falsify the trust and reputation assessment of the reviewed products and services in e-commerce applications. As a result, this falsification would impact customers' trust with regards to the purchase decision-making. Moreover, human users have specific reasons for deliberately skewing their comments and they can change their intentional behavior over time and according to changing circumstances that impact the product's quality and the customer's interests as well. Besides, customers can also be discriminative against particular service providers while cooperating with others.

In order to distinguish honest reputation information from dishonest one, we need a robust TRS that applies intelligent detection algorithms, either supervised, unsupervised or semi-supervised. These algorithms aim to analyze the trustworthiness of the reputation information provided in the form of reviews, recommendations and numeric ratings. In fact, a meticulous trustworthiness analysis of the shared information achieved by TRS, would certainly increase the system's robustness and resistance against fraudulent reviewers. Moreover, the underlying computational goal of TRS is to generate a trustful evaluation of the reputation of a product or a service in e-commerce. Indeed, TRS incarnate a combination of two dependant systems: Reputation systems which is generally based on the rating process, where entities rate each other after the achievement of a transaction [1]. Therefore, an algorithm is implemented in order to use the aggregated ratings about a specific entity to generate somehow a reputation score. On the other hand, Trust systems generate trust measures and scores according to the analysis of combined paths and networks of trust relationships between transacting peers [1, 2]. Reputation scores and trust measures can be combined and derived from a TRS and thereby assist e-commerce users in their transaction and purchase decision making process.

Additionally, TRS analysis must focus on the review provided in the form of text sentence and the numeric score, rating or star rating proposed by the e-commerce strategy of rating, since it disseminate important latent and detailed reputation opinion on specific features of the reviewed product. Relying on Natural Language Processing (NLP) techniques inspired by intuitive human intelligent reflexes and processing, we believe that it is possible to build robust TRS able to create a trusted reputation management network that supports e-commerce applications, boosts their ability performances and reaches the intended business value.

Accordingly, we need to analyze the maturity of current TRS in order to extract their weaknesses and strengths. At this purpose, we consider that a TRS is mature, if a number of credibility aspects and criteria are verified by the System's approach. To gather these trustworthiness aspects, we have focused on more recent surveys on the TRS as well as novel TRS works such as [1, 3, 4] published in 2007, [5, 6, 7] in 2010, [8, 9, 10] in 2013, [11, 12, 13] in 2013, and [14, 15, 16] in 2015. The most relevant trustworthiness aspects that are the most used in the state of the art of the TRS are related to the reviewing and the rating process adopted in the TRS analysis.

Additionally, the selection of recommenders is also considered an interesting trustworthiness criterion that can explicitly impact the trustworthiness and reputation evaluation. In addition, the interpretation and the reasoning on the gathered information remains an indispensible trustworthiness criterion that explicitly acts on the computational approach to calculate the reputation value of the reviewer and the topic reviewed as well. At this purpose, we have adapted these three trustworthiness aspects to our purpose in order to establish a comparative study of a selection of the most popular and studied TRS. Besides, we have extended the comparison framework by adding other trustworthiness aspects. The first one is related to the computation approach to calculate and evaluate the trust and reputation. In fact, there are different computational approaches that can be either based on probabilistic methods to calculate the trust value, or on the Bayesian approach, or on fuzzy rules, statistical methods that derives a percentage representing the trust and reputation estimation, or Dirichlet reputation systems [3], etc... This aspect aims to present and describe reputation computation engines and approaches employed and/or enhanced to compute the reputation and trust value. Basing on this credibility aspect, we can analyze the trust and reputation evaluation results and then critically evaluate the reached accuracy-level of each TRS. In fact, the second trustworthiness aspect is the evaluation of the effectiveness of each TRS involved in the comparative study. Additionally, since text reviews are important in the TRS analysis, we compare the selected TRS in terms of their reliability on opinion or sentiment mining techniques especially for extracting the intended sentiment orientation from text reviews.

In this paper, we present in depth comparative study of the most popular, used and analyzed TRS according to the trust and reputation literature and in terms of specific trustworthiness criteria. We start introducing the chosen TRS for the comparative study. We then compare the TRS in term of the selected trustworthiness criteria. This survey aims to analyze and evaluate their maturity and effectiveness, especially regarding the computational approach employed to generate reputation and trust values. Additionally, we provide a summary table of the compared TRS within a detailed selection of trustworthiness criteria.

## II Overview of the main TRS

When considering online trust and reputation systems, eBay is both well researched and much written about [17, 18, 19, 20, 21]. eBay is an online auction site, allowing sellers and buyers to trade goods through an auction process. The commercial eBay TRS stores users' ratings linked to their profiles and related transactions, but leaves the credibility analysis to the human user.

Unitec is also clearly directed towards a human user, but it performs automated credibility analysis as well. The content of a recommendation is not fixed, and the system could handle product recommendations with the same algorithm as well [1]. A number of TRS are designed for e-commerce applications such as FuzzyTrust [24, 49] and REGRET [25, 51] which are both designed for multi-agent marketplaces, but they apply different approaches to reputation estimation than ebay's TRS [1]. In fact, in FuzzyTrust, local trust scores are generated through fuzzy inference and aggregated to global reputation values. However, REGRET is a decentralized trust and reputation model designed for more complex e-commerce environments where various types of agents with different social relationships play important roles. With the help of a social structure called Sociogram, it is able to model the social relationships such as cooperation, competition and trade in a graph where the nodes represent the participants and the edges denote the nature of their relationship.

In other words, REGRET reputation system is designed to operate within an electronic marketplace settings relying on multiple contextual attributes to classify information as coming from an individual, social, or an ontological dimension. The individual dimension considers information directly gathered from interactions between two entities. The information is fine grained and often related to the frequency of overcharging, late delivery and quality of the transaction [7, 25].

For cooperative applications on the Internet, NICE is a good TRS example. Trustors are given signed receipts of successful transactions, "cookies", as a sign of some trust. These can be used to link actors into a weighted trust chain.

For peer-to-peer ecommerce communities, PeerTrust [28, 15, 39] is a coherent dynamic trust model with unique characteristics tailored. This TRS uses a structured peer-to-peer (P2P) overlay network to host a distributed implementation of their transaction-based feedback system [46]. The simulation used to demonstrate PeerTrust utilizes P-Grid to distribute feedback scores. The system incorporates a combination of fundamental reputation sources, such as direct feedback, and the quantity of transactions performed, while weighting feedback with credibility [1, 7].

For this purpose, PeerTrust defines the personalized similarity measures, which compute feedback similarity rate between the evaluating peer and opinion providers over a common set of peers with whom they have had previous interaction. In fact, this TRS attempts to create trustworthy peers that consistently act honestly as a role of feedback provider, and do not become affected by malicious intentions such as jealousy and negative competitive attitude. Besides, this model assumes that the trust metric can be alternatively served as a credibility measure under certain circumstances [1, 7, 43].

MDNT, Managing Trust and MLE [27, 42, 21, 22] to-peer community environments, which cover both multi-agent marketplaces and cooperative applications. In fact, they can host a multitude of activities especially in distributed applications such as information exchange (file transfers) and transactions in e-commerce applications [1, 7]. Furthermore, their approaches to compute reputation are varying as well.

As previously discussed, file transfers and sharing is a frequent activity in internet requiring security and trustworthiness verification. Indeed, EigenTrust [44] is a reputation system for peer-to-peer file sharing. It relies specifically on a global, shared reviews and recommendations concerning reputation. Credibility analysis is used in calculating each global reputation value [1].

However, for a wide application of open systems such as Grid service that are not restricted to one activity case, TRAVOS [31, 52] (Trust and Reputation model for Agent-based Virtual Organizations) is developed to ensure high quality interaction between the participants. It exploits two information sources to assess the trustworthiness of the participants: Direct Interaction and Witness Observation. To derive trust, this model relies greatly on its direct experiences and refuses to combine others' opinions unless they are really required [43]. In fact, TRAVOS aims to ensure good interactions between self-interested software agents in large scale open systems, such as the Grid. The agents provide interchangeable services, and reputation information is used to choose the most trustworthy partner. Its reputation expression is based on Beta probability distributions [1].

For Multi-agent systems MAS, Fire [33, 51] is designed to handle the bootstrapping problem of newcomers and filter out inaccurate reputation information. In addition, this TRS attempts to differentiate between dishonest and mistaken agents and provide compound reliability measures by employing a multi-criterion rating system.

Moreover, the proposed TRS by Yu & Singh [35] is also suitable for MAS. The system proposes novel trust and referral network able to detect three models of deceptions. It provides credibility measures pertaining to each model and it is adapted to differentiate between agents having bad reputation or no reputation using DempsterShafter theory of evidence. As a result, both Fire and the TRS by yu & singh support dynamism in open MAS [7, 25, 27].

A key benefit of the social dimension is that it allows new and unproven entities to bootstrap their trustworthiness by belonging to reputable groups. Alternatively, because the entire group's reputation is associated with the behavior of its members, it is pertinent for a group's members to moderate the behavior of those associated with them.

Concerning open dynamic environments, BRS [34, 52] remains a very suitable TRS that supports binomial and multinomial rating models and also addresses bootstrapping problem by considering the quality of community in the marketplace. The TRS provides iterated filtering algorithm which can effectively reveal deceptive intentions if the majority of participants act honestly. BRS utilizes longevity factor to discount ratings as time progress, enable participants as buyers and sellers to adaptively change their behavior in order to increase their own benefits [7].

To summarize the overview of TRS selected for the comparative study, we propose the Table III.1 bellow that is an extension of the state of the art presented in [1] representing each TRS, its pertaining domain of application and the different actors involved in the Reviewing and Rating process. In fact, we have updated the related references and added the related information of the following TRS: Fire, BRS, TRAVOS and the TRS proposed by Yu and Singh.

TABLE I
Selected TRS and their application area

| TRS Name | References | Application Domain | Actors |
|---|---|---|---|
| eBay | [55] 2006 [46] 2010 [47] 2015 | electronic Marketplace | human |
| Unitec | [38] 2003 [48] 2014 | generic framework | human |
| FuzzyTrust | [24] 2005 [49] 2012 | multi-agent marketplace | agents |
| REGRET | [25] 2002 [51] 2015 | multi-agent marketplace | agents |
| NICE | [26] 2003 [50] 2012 | cooperative applications | agents |
| MDNT | [27] 2004 [43] 2014 | online communities | agents |
| PeerTrust | [28] 2004 [15] 2015 | online communities | agents |
| Managing Trust | [28] 2001 [51] 2015 | online communities | Agents |
| MLE | [28] 2004 [53] 2014 | online communities | agents |
| EigenTrust | [31] 2003 [54] 2015 | file sharing | agents |

| TRAVOS | [32] 2006<br>[52] 2014 | Grid services | Agents |
| FIRE | [33] 2006<br>[51] 2015 | Multi-agent Systems | Agents |
| BRS | [34] 2002<br>[52] 2014 | Multi-agent marketplace | human |
| TRS by yu & singh | [35] 2003<br>[43] 2014 | Multi-agent Systems | Agents |

In this section, we have presented a brief overview of the most used and researched trust and reputation systems in their different areas of application such as in e-commerce communities, multi-agent systems, Grid services, cooperative applications, peer-to-peer marketplaces...etc. Each one of the presented TRS uses different Reputation evaluation approach according to the effectiveness perspectives. In fact, a number of trustworthiness aspects must be verified in order to analyze the robustness and effectiveness of the TRS. In the coming sections, we will first discuss the taxonomy of the selected credibility aspects and then compare TRS within a critical study using aspects of trustworthiness.

### III  Comparing trustworthiness criteria of the main TRS

In order to properly interpret trust and reputation as measures of reliability and computational concepts, it is necessary to conduct a meticulous comparative study of well-known TRS that involves a range of trustworthiness criteria to be accurately analyzed. Accordingly, we compare various trust and reputation systems in terms of the aforementioned criteria : 1) the strategy adopted to create a review or a recommendation, 2) the process of reviewers selection, 3) the interpretation analysis applied to gather the information, 4) the computational approach applied to calculate the reputation value of the target, as well as 5) the accuracy-level of the trust and reputation evaluation.

#### A.  Reviewing and Rating Strategy criterion

The reviewing and rating process consists of creating and sharing a recommendation and leaving a reputation rating related to an agent, a peer, a reviewer and in general a target topic of review. In order to study this credibility aspect, we present and discuss the content of a review, including how the conveyed experience is expressed and what kind of metadata is provided to support the analysis. In fact, when creating a review, the reviewer generally provides the experience information that generates a new statement to communicate to other entities interested in the same target topic. In general, negative, positive and neutral experience information are provided and used in all the studied TRS except Managing Trust, where only negative reviews are stored.

Storing different experiences and recommendations helps create a valuable distinction and richer analysis regarding the reputation of the target. Taking into account only one category of reviews will restrict the reputation estimation to this category. As a result, the reputation computation will not be transparent but discriminative relying on this kind of approach. Besides, a recommendation can be single rating in the numeric form or star form, created after a transaction with the trustee in order to evaluate its quality or without considering the achievement of the transaction. The recommendation can be also an opinion about the trustee formed by the aggregation of individual ratings. Aggregation saves traffic, providing better scalability, but reducing transparency [1]. Recommendation can be also in the form of textual review expressing opinions, sentiments and experiences. At the end of a transaction, both dealing parties in the transaction can exchange or leave feedback for each other. An achieved transaction is not often a necessary condition to review a product or a service. However, reviewing and rating after an accomplished transaction has more sense and can generate more valuable information but not necessary a trustful one. This kind of Rating and reviewing strategy allows potential future entities to examine the reliability of any target party that has been involved in a previous transaction either the buyer, the seller or the product. Hence, recommendation can take several forms: it can be a single overall rating (Positive, Neutral and Negative), a series of star ratings or a textual review expressed by the recommender. The text review often provides further information about the actual quality of the item, the target topic in general and any problems encountered. we will describe hereafter the Rating and Reviewing strategy adopted by each trust and reputation system presented below.

Starting with the auctions' site eBay [17, 18] its reputation system stores users' ratings linked to their profiles and related transactions, but leaves the credibility analysis to the human user. Their rating strategy relies on Feedbacks scores and stars important to evaluate the reputation of eBay community.

At the end of a transaction, both parties involved in the exchange leave feedback for each other. This allows potential future parties to examine the reliability of any target party that has had a previous interaction. The reputation system documentation [19, 20, 21] assumes that the feedback is left in the form of a single overall rating (Good, Neutral and Negative), a series of numerical ratings (for the following facets: Accuracy, Communication, Shipping Time and Shipping Charges) and a comment. The comment often provides further information about the actual quality of the item, shipping or any problems encountered.

The Feedback score is one of the most important pieces of a Feedback Profile. The Feedback score is the number in parentheses next to a member's username, and is also located at the top of the Feedback Profile. Next to the Feedback score, a star rating is also available. The number of positive, negative, and neutral Feedback ratings a member has received over time, are part of the Feedback score. In most cases, the Feedback score represents +1 point for each positive rating, No points for each neutral rating and -1 point for each negative rating.

A Feedback score of at least 10 earns you a yellow star (  ). The higher the Feedback score, the more positive ratings a member has received. As your Feedback score increases, your star will change color, all the way to a silver shooting star (  ) for a score above 1,000,000! You can view a member's Feedback Profile by clicking the number in parentheses next to the username. If you know the member's username, you can also locate the member's Feedback Profile from the Find Feedback page. Along with a member's Feedback score, the Feedback Profile page also shows the percentage of positive ratings left for a member, recent Feedback ratings, recent comments, any bid retractions, and detailed seller ratings.

As seen below, eBay delegates the credibility analysis to the human user, who can to a degree evaluate the response based on its arguments. However, this delegation may sabotage the users' reputation if comments and rates are falsified for many convincing reasons. In decentralized reputation systems, an actor may never know what kind of statements are made of its behavior. The target's comments are also more challenging and effective to include in an automated credibility analysis. For most systems, the main content of a recommendation is a single numeric value, but the scales used vary between discrete values and real numbers in a range.

In fact, most of the observed ratings on eBay are surprisingly positive. Buyers provide ratings about sellers 51.7% of the time, and sellers provide ratings about buyers 60.6% of the time according to the study of 2002 [37]. In the same study, of all ratings provided, less than 1% is negative, less than 0.5% is neutral and about 22% is positive. It was also found that there is a high correlation between buyer and seller ratings, suggesting that there is a degree of reciprocation of positive ratings and retaliation of negative ratings. This is problematic if obtaining honest and fair ratings is a goal, and a possible remedy could be to not let sellers rate buyers. More details of this problematic are provided in section II.2.2.

However NICE and Unitec TRS do not specify the aggregation method and researches mention that a policy is in place in order to evaluate reputation [38, 48, 26]. Furthermore, it is mentioned that Unitec recommendation is a generic wrapper around any form of value, and data aggregation is based on a local policy that fits the format [1]. In fact, the aggregation policy can be shared among all peers in a network, or locally defined by each peer. Since the system is not tied to a single policy, some generality aspect is provided in order to allow the system to be flexible with a broader application range. However, the user dealing with an application may be limited because of the lack of a global agreement. For instance, sub-networks with separate and independent policies are technically interoperable, but not necessarily holding a semantic relying values. Other TRS rely on experience behaviors and information in their adopted reviewing process. For instance, in REGRET, an opinion is represented as the weighted average of single experience ratings, with more weight given to newer experiences. Moreover, TRAVOS and EigenTrust keep counters of positive and negative experiences, and use them to produce the aggregated opinion. TRAVOS simply presents both counters, while EigenTrust calculates their difference and normalizes the value between the range [0,1]. In both systems, a good experience is defined basing on the agreement that a "trustee" provides the service promised towards the "trustor" entity and there is no consideration of single experiences [1, 6, 24]. However in MDNT, the opinion presented derives a reputation value basing on local experiences and the CCCI metric [45]. In fact, the CCCI metric considers the fulfillment of certain predictable and expectable interactions and transactions the trustor has the intention to achieve, and assigns them weights according to their importance and whether they have been completely and clearly fulfilled by the trustee [27, 43] .

In Managing Trust, only negative recommendations are considered and then stored, and the value of a recommendation is not discussed nor used in the analysis. In addition to the value, the recommendation contains various metadata items. In fact, Unitec, NICE, PeerTrust, Managing Trust and MLE apply mediators that provide meta recommendation. Besides, recommendation time or age is one of the most important factors involved in the reputation analysis. However, not all TRS involve the time information in their reputation analysis [28, 51]. On one hand and unlike these TRS, EigenTrust, REGRET and TRAVOS store recommendations within their time information, which is automatically generated when a recommendation is provided. Consequently, recommendations provided at approximately the same time would have at a pre degree the same weight regarding the time or age factor [25, 50, 6, 24]. On the other hand, FuzzyTrust uses both information about transactions and the recommender's track record as a recommendation. In fact, FuzzyTrust applies transaction times along with other factors to estimate weights for recommendations when calculating the reputation value [1, 24]. However, MLE suggests a time stamp to be included in the storing process of the recommendation as a part of a key constraint, but it is not involved in the reputation evaluation [28, 53. In addition, the privacy of the provided opinion is considered another trustworthiness context factor that can be incorporated in the weighting process. In fact, among the overarching purposes of a TRS is to propagate trust and reputation among online communities by sharing reviews and ratings, either after executing the reputation analysis algorithm or before as an unprocessed information such as in ebay site. Furthermore, it is widely agreed that reviews or recommendations are one of the foundations of TRS since they represent their potential conveyed data source and input. Otherwise, insufficient amount of reviews' input disable TRS to perform effectively [1].

In the meantime, recommendation privacy seems to be a neglected or inconvenient weighting factor in most of TRS reputation analysis. In general, this factor is related to the amount of subjective experiences information provided in the recommendation. In REGRET and Unitec, the recommendation is provided along with the subjective or objective experience privacy of the recommender. Besides in TRAVOS, a privacy measure is implicitly incorporated taking into account the number of good and bad experiences [38, 48, 25, 24]. Additionally, Managing Trust and MLE include the recommender's digital signature as a privacy measure that guarantees the recommendation integrity [28, 50, 22].

However, PeerTrust emphasizes that the aggregation of recommendations, which are only based on the credibility of their recommendations cannot efficiently reflect the trustworthiness of the agents. Thus, other various aspects of the review such as its size, time and category remain indispensible recommendation context factors that are explicitly incorporated in the Reputation analysis. These factors are also considered trustworthiness measures that are included in the detection of the reviewers' intentions and potential fraudulent transactions.

In the context of TRS extending and adapting existing reputation system such as the Bayesian reputation system, Jøsang et al. in [34] have proposed the flexible and adaptive Bayesian Reputation System (BRS). This TRS supports both binomial and multinomial rating models to allow rating provision happen in different levels of precision well suited for open dynamic environment. Theoretically, multinomial BRS is based on computing reputation scores by statistically updating the Dirichlet Probability Density Function [3]. In binomial context which is based on Beta distribution, the BRS agents can only provide binary ratings for the others.

Besides, Yu and Singh propose a novel trust network which intends to locate the most appropriate witnesses in a multi-agent system. In this model, each agent is surrounded by a number of acquaintances among whose subsets can be neighbors. When a requesting agent wants to evaluate the trustworthiness of a particular agent, he needs to send a query to the neighbors of that agent asking them for their perception and opinion regarding the target agent. Unless the neighbors have not established any direct experiences with that agent, they respond by their witnessing experiences and proofs. Otherwise, they will reply by returning a series of referrals [7]. The number of referrals is limited by the branching factor and depth Limit parameters, so as to limit the effort expended in pursuing referrals. This process successfully terminates if an adequate number of ratings are received. This TRS encounters failures when the depth Limit is reached and neither ratings nor referrals are gathered [35].

Moreover, FIRE employs a multi-criteria rating system in which the participants rate the performance of Service Providers SP based on predefined criteria, rather than providing an overall opinion. However, it does not provide techniques to perform context diversity checking, whereupon the context and criteria similarity rates are not calculated. Thereby, adopting this rating process makes the TRS unable to elicit the preferences of relying agents in order to predict the trust value of a particular SP more precisely [7, 25].

According to the previous analysis, we conclude that rating and reviewing processes remain a distinguishing criteria in the comparative study of different TRS. In fact, These systems apply various rating techniques in order to predict and obtain a trustful and reliable reputation estimation of the target agent or topic. Many rating processes and techniques have been presented in this section such as multi-criterion rating process, binomial and multinomial rating, numeric rating, text and statistical ratings, basic probabilistic methods, Bayesian probabilistic methods as well as enhanced ones, Dempster Shafter theory of evidence and Dirichlet methods. Thereby, the characteristics of each rating techniques will considerably impact the process of selecting recommenders and reviewers in addition to the reputation computation approach to be effectively performed. Thus, the approach applied to select potential agents and reviews is the next discussed credibility aspect incorporated in the comparative study.

*B. Reviewers' Selection process criterion*

Various factors are taken into account in the process of selecting or grouping reviewers, recommenders and agents involved in online transactions. Social context features, transaction context criteria, relationship with the trustee or the trustor, a membership feature, history transaction background, history positive and negative interactions with a peer, trust and reputation value, security requirements such as digital signature, authentication security...etc are a discerning criteria carefully analyzed while selecting reviewers.

Starting with REGRET, the technique applied to select raters consists of grouping peers according to their social context, such as their relationship with the trustee, and then chooses the most representative member of each group to give recommendations by fuzzy rules [7]. In fact, REGRET includes the individual dimension or subjective reputation which relies on the direct impressions of an agent received from Service provider (SP) and prioritizes its direct experiences according to their recency. The social dimension is also incorporated in case the agent has newly joined the environment. This dimension is itself divided into three specialized types of reputation depending on the information sources. First, witness reputation which calculates reputation based on the information coming from the witnesses adjacent to this agent. Here, adjacency represents a sort of an existing relationship between two agents. Second, neighborhood reputation that measures the reputation of individuals who are neighbors with the agent being evaluated by considering their social relationships and third, system reputation which assesses the trustworthiness of SP based on the general role that it plays in the sociogram [7, 25].

In peer-to-peer online communities, it is commonly frequent that peers rate each others' interactions, positive and negative experiences as well as performances. Therefore, these votes and ratings are incorporated in the Reputation and trust computation algorithm. To elicit peers participations in the rating process, they are rewarded discounts, trust and reputation rating levels until reaching the top level...etc. For instance PeerTrust, Eigentrust, BRS, TRAVOS and the TRS designed by Yu and Singh are systems that motivate peers to rate each other in different ways and for different purposes.

In PeerTrust, recommendations are saved in a distributed manner, searchable by the trustee's identification. The mediator repeats the set of signed recommendations he is responsible of their storage. Therefore, to encourage participants' cooperation, PeerTrust embeds a reward function, called the community context factor, into the trust metric to encourage peers to persistently provide votes about others' performance. The dynamic and distributed nature of peer-to-peer systems necessitates an optimized and adaptive design of the peer location approach. To concretize this objective, this model provides each peer with a trust manager and a data locator engine which are responsible for feedback submission and retrieval aside from trust evaluation over the underlying network [7].

Besides, EigenTrust [44] allows entities to decide which peers they will trust toward files transfer and downloading. This TRS is completely decentralized, and relies on a distributed hash table overlay. Besides, the reputation framework builds a history background of other entities transactions. In fact, each agent keeps a personal history for other peers, represented by the sum of positive and negative interactions they have experienced with them [1].

Concerning BRS [34, 52] the system enables participants as buyers and sellers to adaptively change their behavior in order to increase their own benefits. Moreover, agents are allowed to rate other peers within any level from a set of predefined ratings levels.

It is noteworthy to mention that reviewers and peers experiences are somehow taken into account in the trust metric of each TRS. As previously discussed in REGRET, EigenTrust, PeerTrust and BRS, raw experiences were integrated in the calculus of reputation value without a serious preprocessing task that would reveal their ineffectiveness or untrustworthiness. To overcome this limitation, TRAVOS provides a confidence metric that determines whether the personal experiences are sufficient to make an acceptable judgment with respect to a particular service provider or not. If not, it disseminates queries to obtain additional observations from other witnesses who claim to have had previous interaction with that certain service provider. Specifically, this TRS utilizes a single rating system that summarizes interactions in a single variable, which indicates an overall performance. Here, witnesses share the history of their interactions in a tuple which contains the frequency of successful and unsuccessful interaction results. Moreover, in order to deal with inaccurate reputation providers, TRAVOS takes advantage of an exogenous approach. According to this approach, instead of calculating the reliability of the provided recommendation based on its deviation from mainstream opinions, it calculates the probability that a particular correspondent provides accurate reports given its past opinions and proportionally adjusts the influence of its current observations afterwards [31, 52, 7]. This approach has demonstrated a high accuracy-level surpassing available approaches in the same context of involving peers experiences and mutual ratings in the reputation computation algorithm.

Another enhanced approach regarding the trustworthiness of provided experiences is adopted by Yu and Singh TRS. In fact, the system applies various distinctive features which surpass other available models in some contexts. These features consist of distinguishing agents' interactions, either they directly established or not. Furthermore, in case of absent ratings, an alternative method must be agreed. In fact, this model of reputation management exploits two information components. The first one contains the agent's local belief created as a result of its direct interaction with other agents. The second one includes the testimonies of third-parties that can be beneficial in the absence of local ratings. The requesting agent combines a local belief in conjunction with third-party testimonies to achieve a more accurate estimation the total belief regarding the trustworthiness of a particular agent. Additionally, each individual agent maintains a two-dimensional model of each acquaintance. The first dimension indicates their ability to act in a trustworthy manner, which is called expertise and the other one signifies their sociability in referring to suitable trustworthy agents. Depending on their competency in fulfilling either of the above-mentioned qualities, acquaintance models are modified to reflect their actual performance to be used in future interactions [35, 7].

In addition to these relevant factors, it is necessary to design a robust TRS able to detect and encounter malicious agents and their interactions in different systems, especially in open multi-agent systems MAS involving dynamic characteristics. The decentralized FIRE framework is developed to deal with this kind of environment. This TRS is engaged in the detection of malicious third-parties who provide misleading reports. In fact, FIRE builds a credibility model to assess the honesty of the revealing reports and then filters out the lying and untruthful reporters. Besides in FuzzyTrust, peers' roles are established in the network in order to evaluate peers' reputation. In fact, recommenders are chosen through global weighting based on their number of performed transactions and local trust score. The weight is then compared to a threshold that is set based on the peer's role in the network a super peer with a high number of transactions has a higher threshold than a regular peer, for load balancing reasons. The influence of peers with many transactions is thereby somewhat reduced [7, 1, 24].

In all previous TRS, agents and peers are identified in order to estimate the credibility based on the source information. However, Managing Trust is the only system in this comparative study that allows reviewers to remain anonymous. Indeed, recommenders remain anonymous, but the credibility of the mediators responsible of their recommendations storage is estimated based on their transaction history. Otherwise, a complete anonymity is not a reliable criteria to estimate the credibility of the source of information. However, approaches can be applied to overcome this limitation such as involving trusted third party mediators to provide the privacy desired from recommenders.

The anonymity is not the only neglected issue in TRS reviewers' selection process. Misbehaving users that well understand the system's analysis, can take advantage from their knowledge to promote their reputation and turn the balance on their side. eBay's TRS is one of the systems that suffers from misbehaving users. In studies realized on eBay's reputation system in 2015 presented in the following table III.3.2 (1), the percentage of time that sellers leave positive, negative or neutral, or no feedback is surprising. Another criterion included in the study is the initiative of rating that indicates which one is moving first for rate is it the seller or the buyer. This aspect can reveal a priori some misbehaving engagement and plot between the seller and the buyer. The data is divided into two segments. The first segment is when the seller is moving first, before the buyer leaves feedback. In this segment, the authors of [14] also divide the data conditionally on the buyer's response after the seller's feedback. The second segment is when the buyer has left feedback before the seller. The seller's response is reported conditionally on the buyer's action.

Figure 1. Sellers' Feedback on eBay's market [14]

|  | Feedback Left by Sellers | | |
|---|---|---|---|
|  | Positive | Negative or Neutral | No Feedback |
| *A. Seller Moves First* | | | |
| All | 98.83% | 1.17% | – |
| Positive | 99.95% | 0.05% | – |
| Negative or Neutral | 91.94% | 8.06% | – |
| No Feedback | 96.7% | 3.3% | – |
| *B. Buyer Moves First* | | | |
| Positive | 88.47% | 0.04% | 10.49% |
| Negative or Neutral | 5% | 37% | 58% |

Additionally, a new study [36] concludes that some eBay users are artificially boosting their reputations on the Internet auction Web site by selling items for practically nothing in exchange of positive feedback from the buyer. Sellers with good reputations can seek higher prices on items they sell, according to the study out of the University of California at Berkeley's Haas School of Business [36]. In fact, under eBay's reputation system, buyers and sellers can submit feedback to each other after a transaction. Some merchants are selling items at minimal prices, such as 1 cent. They then hope that grateful buyers will give them positive feedback, ask for it or offer to provide positive feedback in exchange, according to professor John Morgan, who co-authored the study. Authors also states that merchants were selling written compliments in exchange of positive feedback.

Basing on this report concerning eBay's misbehaving sellers and buyers, we can conclude that this kind of attitude must be prohibited in order to increase the effectiveness of the reputation system and to better reach the trustworthiness and credibility intended.

Basing on the previous analysis, we notice that a number of context criteria is taken into account during the selection of peers, agents and reviewers. For instance, social context features, history of past transactions and interactions with other agents, experiences background remain the most reliable features of selecting reviewers and agents in the reputation computation analysis. However, some features must be carefully involved in the process incorporating some pre-enhancement. In fact, personal experiences report and sharing may be insufficient to make an acceptable judgment toward a particular agent. In that case, additional analysis is required and involving witnesses testimonies can effectively respond to this purpose.

Additionally, reviewers identification transparency is necessary in order to estimate the trustworthiness of the source of the provided information. Once shared, this information undergoes a number of preprocessing tasks in order to verify its integrity and to be incorporated in the reputation evaluation process. As a result, an interpretation analysis of the gathered information remains an interesting credibility criteria in the current comparative study of TRS.

*C. Interpreting the provided information criterion*

Once reputation information is collected, it must be aggregated into a suitable format to support trust decision analysis. For this purpose, the third trustworthiness criteria evaluates the analysis applied on the gathered information. In the literature, there are a selection of factors that are employed to interpret the reputation estimation. These factors are tightly related to participants' behaviors and experiences, the availability of information, the variability of the individual ratings, the confidence of the recommender and the social relationships. However, insufficient or absent direct experiences and interactions unconstructively assess the reputation evaluation. to overcome this limitation, witnesses observations and opinions regarding participants are involved in the interpretation process as well as in the reputation computation algorithm. Additionally, history transactions factor and time information are widely employed by the TRS in order to update the trustworthiness estimation according to their recency. In this section, we compare TRS interpretation of the gathered information according to these factors. For this purpose, we first briefly present the aggregation of recommendations into a reputation estimation in order to interpret the final reputation evaluation. We then examine critical issues that are marginalized by TRS and that automatically lead to a falsified interpretation of the reputation estimation. We additionally discuss the category of the information history stored and employed by some systems, in order to adjust the system context to their application area and social setting.

Dellarocas in [40] states that the aim all reputation systems especially eBay's feedback mechanism, is to ''generate sufficient trust among buyers to persuade them to assume the risk of transacting with complete strangers''. Despotovic and Aberer [42] discuss ''reducing the opportunism'' and vulnerability of both the trustor and the trustee. In [41] it is agreed that using a trust and reputation system, a party may examine the history of another and decide that it will trust and interact with the other party. This decision is often called a ''trust decision''.

However, in developing centralized TRS for online auction applications like eBay, the availability of information sources is not an issue. More explicitly, since the reputation computation engine is taken over by the central authority, it can ubiquitously access the pool of information upon demand. Furthermore, the ballot box stuffing and correlated evidence problem is out of question in a centralized system [7]. However, the main concern of such systems is the absence of an effective rating system which could articulately assess the delivered services analogous with real-life experienced judgment. To support this perspective, [22] claims that provided feedback in eBay is unrealistically positive such that service providers receive negative rating only 1% of the time, which indicates the incapability of such systems in providing informative ratings.

Interpreting results in terms of credibility or reliability is the main purpose of reputation analysis engines. In fact, threshold-based values, rank-based and fuzzy-rule estimations are commonly attributed to evaluate the source of information, peers' ratings and performances, social relationship between reviewers, the privacy and confidence of the reviewer etc. For instance, REGRET calculates a reliability value for each type of reputation, based on a variety of factors such as the available information, the variability of the individual ratings, the confidence of the recommender and the social relationships. It also estimates a reliability measure for the final estimated reputation. In order to be able to calculate social reputation, REGRET must first identify appropriate witnesses in the e-commerce environment. For this purpose, the TRS applies graph theory techniques to the sociogram to locate the most appropriate witnesses and examines their social relationships with the agent being evaluated. The ontological dimension is used by REGRET in order to add the possibility of combining different aspects of reputation to calculate a complex one. In addition to the reputation value, REGRET comes with a reliability measurement which reflects the confidence level of the produced reputation value [7, 25].

Besides, PeerTrust suggests calculating a personal similarity measure as an alternative for estimating recommender credibility. Similarity measures are commonly used in product recommender systems, where credibility cannot be measured objectively. PeerTrust also proposes the compilation of a confidence value associated with the reputation estimate, but instead of a specific measure, it presents a tentative suggestion for the value to reflect the number of transactions and the standard deviation of the recommendations depending on different communities [28, 15, 7]. In addition to the peer's behaviors, the credibility of the estimated reputation value depends also on the history of transactional information. Indeed, eBay, PeerTrust and Managing Trust store the full transaction history of agents and their behavior. Besides, EigenTrust uses the resulting reputation value as a probability of selecting the trustee as the peer to transact with. Furthermore, in TRAVOS community, witnesses share the history of their interactions in a tuple which contains the frequency of successful and unsuccessful interaction results. This model merely relies on direct experiences of agents to derive trust. However, if direct experiences of target agents are absent or insufficient, the system combines others' opinions provided form witnesses. For this purpose, the TRS relies on experiences factors that are indicated by Direct Interaction and Witness Observation, in order to assess and interpret the trustworthiness of participants.

Nevertheless, initial values of trustworthiness estimation must be applied for newcomers, who obviously have no history transactions background yet. In fact, initial trustworthiness estimation can be interpreted either as no reputation or bad reputation depending on their threshold-based values. According to the Dempster's theory, lack of belief does not necessarily imply disbelief in the Yu and Singh's system. Thus, instead of assuming total disbelief as initial value for newcomers, it is replaced by a state of uncertainty. In other words, with the help of the theory of evidence, Yu and Singh's model is able to differentiate between having a bad reputation and no reputation at all. Moreover, to predict total belief, it utilizes Dempster's rule of combination as an aggregation method which combines evidences to compute new belief value. In addition, this model describes a variant of the Weighted Majority Algorithm (WMA) in order to better tune the weight of advisers for the purpose of deception detection after actual successful or unsuccessful interactions [7, 27].

Besides, time information is also an interpreting factor that helps identify the recency of transactions and agents interactions. This factor is also necessary in the updating process related to agents and their provided recommendations in order to assess their reliability and trustworthiness. BRS is one of the TRS that interprets reputation estimation regarding time information. For this purpose, BRS defines a recursive updating algorithm based on the longevity factor to update the participants' reputation scores in certain time intervals. This recursive algorithm also provides a measure to calculate convergence values for the reputation scores.

According to the comparative study above, we can conclude that there are many reliable factors that help assess the trustworthiness of agents, peers, participants and general target entity or topic. On one hand, some of the proposed factors such as experience behavior and transactions, direct experience, direct interaction seek for evaluating trust according to the available information. In fact, there are two information sources to assess the trustworthiness of the participants according to the experience context: Direct Interaction and Witness Observation. Thus, Witnesses observations and opinions can be exploited in order to deal with the lack and insufficiency of direct experiences and interaction that lead to a mediocre accuracy-level.

Moreover, the more relevant factors are used the more accurate the reputation evaluation will be and assigning higher weights to recent recommendations about the peer's behaviors is considered more likely an indicative and predictive factor of the future behavior. As a result, the information history is involved in the reputation network and can be divided into two categories: transactional history, and the history of accurate recommendations. In addition, time information is also included in the reputation estimation analysis. The insertion of these factors aim to obtain more accuracy by updating the trustworthiness values according to the recency and longevity convergence estimations.

### D. Reputation Evaluation criterion

In the literature, there are several reputation computation engines and approaches that can be broadly categorized in the following categories: deterministic approaches, basic probabilistic approaches, Bayesian probabilistic approaches and fuzzy model. However, many enhanced TRS rely on existing and evolving systems such as Dirichlet reputation systems [3] in order to yield the desired results. In this section, we propose a comparative study of TRS based on their different reputation computation approaches.

On one hand, number of TRS commonly opt for a threshold-based interpretation of the final results. In fact, NICE, MDNT, Managing Trust and MLE rely on threshold values in order to assess the trustworthiness of participants as well as entities. Indeed, if the trustee's reputation value is high enough, the trustor will decide to transact with it. In MDNT for instance, a reputation estimation involves predicting the trustee's behaviour probabilistically, based on experience from a specific time period. MLE also uses a probabilistic approach, and relies on the probability of recommenders to provide incorrect information [1]. Managing Trust allows the trustor to evaluate the trustworthiness of recommenders iteratively when their provided information does not help making the right decision about the target topic. However, eBay and Unitec do not perform the interpretation at all, as they present a report to a human user. In eBay, two interpretations are possible, with thresholds more common. Generally, rank-based selection is only usable when several potential partners can provide a similar service and are therefore interchangeable.

On the other hand, FuzzyTrust, REGRET and PeerTrust do not specify their adopted approach to interpret the final measure, but both threshold and rank-based approaches are possible. However, a fuzzy approach for aggregation is applied by FuzzyTrust, when aggregating personal trust ratings. Only TRAVOS explicitly uses a rank-based approach, where a trustor selects a group of potential partners according to their reputation. That is, TRAVOS considers the current results of all previous interactions with a group of Service provider, where the agent provided similar observations. Then, comparing the variables of their beta distributions, the system measures the degree of accuracy of that specific agent. Thus, truster agent constructs a beta distribution of the rater's current opinion and calculates the relevant expected value $E_r$. It also builds the beta distribution of all the previous outcomes in which the rater has provided similar opinions and estimates its expected value $E_0$ as well [7, 1].

Aside from the social dimension, REGRET applies the ontological dimension, where reputations collected for atomic aspects are combined to construct more complex graph structures in order to derive further insight. Ratings, or impressions, are recorded as a value between positive and negative one. An entity's reputation is then the aggregation of the result of all transactions they have taken part in. When utilizing the ontological dimension, each atomic aspect is calculated using individual and social dimensions, and then combined through a weighted graph for more complex evaluation. The computation of the ontological reputation, $OR_{ij}$ is achieved through Eq. (1) extracted from [43]. The REGRET system also employs a degree of reputation decay, called a forget factor related to the recency of transactions, where only the most recent transactions are considered [7, 42, 25].

Moreover, the rating values of EigenTrust are normalized between 0 and 1. An entity calculates the overall Trust value for another entity, $T_{ij}$, by using personal histories which are obtained from others in on the network. These histories are weighted by the credibility of the reporting entity, as seen in proposed by Ferry Hendrikx et al. [43] in Eq. (3), where $e_{ix}$ denotes a local trust value of entity i for entity x. In fact, the system initially uses the direct experience of others and a local impression of the reporting peer to calculate trust. This approach aims to compute the global trust of i which aggregates the trust evaluation that each peer has provided concerning i and over a time period k. Additionally, α is a constant less than 1 and p is used as a parameter to add trust to new entities in the network [42, 44].

Besides, the PeerTrust TRS introduces two novel trust metrics which are represented by a community context factor and a transaction context factor. The simulation presented by the TRS has entities that generate a binary rating of either zero or one. The parameter $N_i$ represents the total number of transactions, where the entity i has taken part in. Other parameters are involved in the computational equation such as $P_{ij}$ that indicates the other entity involved in the transaction, $S_{ij}$ which is the normalized level of satisfaction i received from peer $P_{ij}$ from the transaction. In addition, $C_r$ denotes the trustworthiness of the feedback received from the entity $P_{ij}$, $TF_{ij}$ represents the adaptive transaction context factor for entity ij toward the transaction, and CF denotes the community context factor for i during a period of time. The normalized weighted factors α and β are respectively the community context factors representing the collective evaluation [42, 20].

In summary, this model customizes a variety of common factors in order to enhance advanced assessment and quantification of peer's trust value in constantly evolving environments. These relevant represent: 1) feedback which is a judgment review of other peers regarding target peer; 2) feedback scope such as the amount of transactions the peer experienced with others; 3) credibility factor for evaluating the honesty of feedback sources 4) transaction context factor such as time and size of transactions which could act as defense mechanism against fraudulent interactions of malicious agents; and 5) community context factor that addresses the feedback incentive problem [7, 20].

Contrarily to PeerTrust which allows peers to text review each other, in BRS multinomial context, agents are allowed to rate other peers within any level, but from a set of predefined rating levels. However, in binomial BRS which is based on Beta distribution, the agents can only provide binary ratings for the others. That is, in multinomial BRS, reputation scores do not only reflect the general quality of service, but they are also able to distinguish between polarized ratings cases and of average ratings ones. Such differences are manifestly not noticeable in binomial ratings, which plainly generates uncertainty and low credibility rate in aggregated reputation score. In addition, this might prohibit the reputation scores to converge to specific values. Furthermore, multinomial BRS allows the input ratings to be provided based on both discrete and continuous measures to reflect a rater's opinion more accurately when required. In order to achieve this goal, it exploits the fuzzy set membership functions to transform continuous ratings into discrete ones, in order to provide compatible inputs for BRS. Both systems use the same principle to compute the expected reputation scores, by combining history interaction records with new ratings [7, 26].

The major concern of the Yu and Singh's TRS regarding the trustworthiness evaluation, is dealing with malicious agents who intentionally propagate propaganda through network to promote their reputation. The TRS proposes three types of deceptions or malicious interactions: complementary, exaggerative positive and exaggerative negative. This classification relies on the behavioral model of the participants in giving ratings. For instance, if agents intentionally give controversial ratings, they may be detected as malicious agents with complementary model of deception. Such agents will lose trustworthiness in the update phase. Similarly, an agent with exaggerative positive or negative tendency of rating seems to be untruthfully acting in the system and will be also loosing from his trustworthiness level. To clarify, the credibility of an agent is reduced according to his dishonesty [7, 27].

In the trustworthiness model of Fire, a threshold of values is defined in order to represent the maximal acceptable differences of the exposed reputation values with the current interaction result. Therefore, any recommender whose inaccuracy exceeds the threshold is labeled as dishonest and is seriously reprimanded by losing his reputation value. An honest but mistaken agent is obviously not excluded from the penalization process. To overcome this exception, a potential issue would be to tune the threshold to a higher value to reduce the probability of falsely classifying the honest witnesses. However, this solution is delaying the process of discarding dishonest agents. To effectively resolve this concern, FIRE develops learning techniques to automatically customize thresholds' values according to the performance deviation of a service provider, relying on direct observations of the evaluator agents [33, 50, 7]. However, there still be a trade-off in distinguishing deceptive agents from mistaken ones.

From the comparative study above, we can conclude that various reputation computation approaches were applied according to the TRS application are, to the reviewing process as well as the reviewers' selection approach. Each of the applied algorithm to derive trust demonstrates its effectiveness in some area, but still suffer from drawbacks inherent in its nature. In fact, the most challenging concern of the applied reputation computation engines is to opt for the right threshold of reputation values and somehow tune them to the most appropriate value in the right moment and the right situation. Moreover, detecting and discarding deceptive reviewers is one the tasks that can plainly judge the TRS effectiveness. For this purpose, several methods of penalizing dishonest agents and rewarding honest one were established, in addition to determinant factors such as social dimension, time information and longevity parameter, ontological dimension, witnesses observations, third-part experience, history transaction background, ...etc. Thus, the implementation of these different factors remains a tough and tedious costly work. However, the studied TRS succeed in the implementation phase of these parameters. Nevertheless, some enhancement could increase the effectiveness of the TRS in order to result a better reputation value. In the following section, we highlight details of the accuracy-level of each TRS.

*E. Effectiveness Evaluation criterion*

Basing on the reports previously discussed that shows eBay's misbehaving sellers and buyers, we can conclude that this kind of attitude must be prohibited in order to increase the effectiveness of the reputation system and to better reach the trustworthiness and credibility intended. On one hand, the effectiveness of eBay's reputation system is deteriorated because of the huge amount of "falsified" and "untrustworthy" positive feedback. On the other hand, provided opinions and ratings require deep automated Trustworthiness analysis in parallel with the human user recommendation and analysis. As a result, eBay's reputation system does not employ any NLP processes in order to analyze text reviews written by reviewers or determinate the conveyed sentiments or intentions. Consequently, this TRS is merely relying on human's interactions and recommendations without any transparency and mutual verification and trustworthiness analysis. Meaning that it is very distant from the intelligence and robustness required in an effective TRS .

Additionally, the TRS effectiveness can be measured according to the system's adjustment to the area of its application. For this purpose, authors of [1] analyzed whether systems were adjusted to their environment of operation, and made a summarizing conclusion that there are two design approaches to be developed, in order to maintain the system's adjustment. The first is related to the reputation computation engine and states that allowing more than one level of reputation calculation is an adjustment to the technical network environment. It adapts the system to the variability of the situation contexts so as to differentiate between very clear or "routine" decisions and situations requiring more careful analysis. The second design relies on addressing the tendency for reciprocity and fear of reprisals in order to deal with the social environment, while involving parameters changeability describing the typical human behavior. Furthermore, human reviewers have a tendency to reciprocate such as to respond to positive by positive and to the negative by the same manner.

In fact, the fear of reprisals can lead to positive recommendations out of fear of negative consequences. Managing Trust protects against reprisals by keeping the original recommender anonymous. The authors describing Unitec also include suggestions for adding some privacy for the recommender [38, 48]. This kind of protection helps avoid the consideration of the environment factors that impact reputation information. In fact, in order to tackle with the variability of context factors inherent in the openness of multi-agent systems, Yu and Singh TRS applies the Dempster-Shafer theory of evidence as an underlying computational framework.

Moreover, several TRS include two or more different levels of reputation calculation based on the information available. Sufficient local experience information can be used alone to calculate a reputation value. Otherwise, If local information alone is insufficient, recommendations are involved in the evaluation analysis as well [1]. MDNT and TRAVOS rely on this approach to derive trust. REGRET has even more levels of fallback, such as calculating the reputation of the social neighborhood of the trustee.

Additionally, a number of TRS such as REGRET, Fire, TRAVOS, BRS and EigenTrust rely on transaction Context factors such as the social dimension, the ontological dimension, history of transactions and changeability and variability of transactions features in order to increase their effectiveness.

In fact, by presenting the social relationship in the form of fuzzy rules, REGRET is able to determine the trustworthiness and credibility of the shared recommendations, by assigning appropriate describing weights to them. For instance, it may declare that if the competition attitude of a witness regarding the target agent is very high, Then its recommended reputation value should be very low. Similar to FIRE, trustworthiness measurement is calculated from a combination of two factors: the number of available recommendations and the variability of the recommendation values. In order to boost in advance the accuracy of the trustworthiness measure, REGRET proposes an intimacy level of interaction which indicates the maximum number of good-quality interactions required to generate a close relationship. As the number of impressions grows, the trustworthiness degree increases until it reaches a certain intimate value. Afterwards, trustworthiness is not affected by the increment of the intimate parameter. It is noteworthy to mention that the value of the intimate factor is dynamically adjustable depending on the frequency of individuals interactions as well as their quality [7, 25].

In addition, FIRE is able to handle potential problems in open multi-agent systems such as the scalability engendered due to the openness nature of the environment as well as the variability and changeability of participant's behaviors and relationships. In order to maintain an effective operation under such circumstances, Fire continuously monitors the performance of components and adopts learning techniques with the purpose of adjusting respective parameters customized to the current situation. It is also a generic model which can be instantiated and applied in a wide range of applications. Moreover, thanks to the variability and diversity of information sources, FIRE is able to effectively derive trust in almost any situation [7, 25].

Furthermore, PeerTrust proposes a novel trust metric that includes a number of parameters to enhance accuracy and reliability of predicted trustworthiness. In order to undetectably continue sabotaging the system, malicious participants aim to maintain their general trust value at a certain level by increasing the transaction volume which falsely dissimulates the impact of their frequent frauds. To alleviate the effect of these malicious attacks, the system combines feedback which is a judgment of other peers regarding target peer and the feedback scope instead of simply aggregating generic feedback values. This approach helps peers circulate their degree of satisfaction by calculating the average amount of successful outcomes that they experienced. Besides, to ensure the reliability of the reputation information, peers are equipped with trustworthiness measures to compute the credible degree of satisfaction. Experimental results demonstrate a high accuracy-level when adopting this behavioral approach [7, 20].

Besides, BRS appears to be a promising method to improve trust evaluation among strangers in an online environment. It adopts an innovative approach which enables trustee agents to evaluate the truthfulness of the ratings provided by reviewers. For this purpose, it applies the endogenous discounting method to exclude such advisers whose probability distribution of ratings significantly deviate from the overall reputation scores of the target agent. That is, it dynamically determines upper and lower bound thresholds, in order to adjust the iterated filtering algorithm's sensitivity tailored to different environmental circumstances. For instance, if the majority of participants are deceptive in the environment, the lower bound would be set to a higher value so as to increase the sensitivity of the BRS, which can lead to the exclusion of more unfair raters. Besides, in order to deal with dynamicity in the participant's behavior, BRS provides a longevity factor which determines the expiry time of the old ratings and gives greater weight to more recent ones. All these considered parameters significantly improve the accuracy level of the system [34, 52].

REGRET, TRAVOS, EigenTrust, MDNT, eBay, PeerTrust, Managing Trust, Nice and FuzzyTrust are TRS that use information about transactions and the recommender's track record, while MLE uses a subset only. In NICE, the task of storing the transactional history is divided between the trustor and the trustee and recommendations expire after a given period of time. However, Unitec does not store history, but a single current opinion value that is updated by each experience item. This is a scalability trade-off in exchange for some lost details and can nevertheless lead to deficient and untruthful results.

Besides, recent transactions are given more weight in reputation estimation in many systems, which is the convincing reason to include time information. FuzzyTrust uses transaction times along with other factors to estimate weights for recommendations so as to compute the reputation value. In one proposed version of reputation estimation metric for PeerTrust, transaction times are used as a part of a transaction context factor that weighs reputation of recommendations. Both REGRET and MDNT use transaction times as well when evaluating the reputation of a review without sharing the time information. In addition, MDNT uses recommendation time information as a recency boosting parameter, in order to update the trustworthiness of the concerned recommender. Nevertheless, eBay does not use time information in the numeric data, but it sorts the text reviews by recency while sharing the times of transactions and recommendations with all users. Besides, NICE uses the recommendation time to prune old recommendations. However, Unitec does not specify its aggregation method nor its credibility calculation, but since it stores the time of recommendation, it could be used in the reputation computation. Managing Trust, MLE, EigenTrust and TRAVOS do not use transaction or recommendation recency information, they therefore consider all experience information identical regarding time criteria. Besides, authors of [7] states that TRAVOS is able to estimate the honesty and accuracy of a current rater's observation. This TRS attempts to decrease the effect of unreliable opinions on a final computed reputation value.

From the evaluated systems, we state that an insufficient amount of gathered information converges to uncertainty, and the credibility of information sources strongly impacts the credibility of the result. Additionally, the consideration of transaction context factors is indispensible in the reputation computation engine in order to derive reliable results. We also emphasize that flexibility in reputation evaluation allows TRS to balance between more accuracy and less complex calculations responding to the scalability of environment [1]. Furthermore, general reciprocity and fear of reprisals are only rarely considered by the trust and reputation systems and can easily skew the intended results.

## IV Summary of the TRS comparative study

In order to summarize the comparative study realized on the main used TRS, we selected the most distinguishing trustworthiness criteria previously described. We also propose to extend some of the trustworthiness criteria in order to derive interesting conclusions from this summarization.

For this purpose, we propose to synthesize the trustworthiness criteria involved in the analysis in six different categories as follows:

-1) Reviewing strategy:

This category involves criteria that recapitulate the rating and reviewing process adopted by a specific TRS. The "text review" column in Table II.3. indicates the possibility of writing a text opinion expressing sentiments, emotions and disseminating recommendations about a target topic. Another interesting criteria is the category of available and stored reviews: whether the system considers both "positive and negative" reviews or only one of the two.

TABLE 2  Summary of the TRS analysis within a set of credibility criteria

| TRS Name | Text review | positive and negative review | Combination of rating and text review | Concordance between rating and text review | Transaction context factors | History Record | time factor | Reputation Value | Reputation value update | malicious agents Control | Third-part witnesses observation | Reviewer's confidence |
|---|---|---|---|---|---|---|---|---|---|---|---|---|
| eBay | yes | only positive | No | No | No | yes | No | statistics, text | No | No | No | yes |
| Unitec | yes | yes | Not mentioned | No | No | No | possibly | configurable | Yes | No | No | yes |
| FuzzyTrust | No | yes | No | No text review | yes | yes | yes | numeric | Yes | yes | No | No |
| REGRET | yes | yes | yes | No | yes | yes | yes | numeric, categorized | Yes | yes | yes | yes |
| NICE | yes | yes | Not mentioned | No | yes | yes | yes | configurable | Yes | No | yes | No |
| MDNT | yes | yes | No | No | yes | yes | yes | numeric, {0,..,6} | Yes | No | No | No |
| PeerTrust | yes | yes | yes | No | yes | yes | yes | numeric, [0,1] | Yes | yes | yes | No |
| Managing Trust | yes | only negative | No | No | No | yes | No | numeric, -1, 0, 1 | Yes | No | yes | yes |
| MLE | No | yes | No | No text review | No | yes | No | probability | Yes | No | yes | yes |
| EigenTrust | yes | yes | No | No | No | yes | No | statistics, text | Yes | yes | No | No |
| Travos | yes | yes | yes | No | No | yes | No | probability | Yes | No | yes | No |
| FIRE | yes | yes | yes | No | yes | yes | possibly | multi-criterion rating | Yes | yes | yes | No |
| BRS | No | yes | No | No text review | yes | yes | yes | binomial and multinomial ratings | Yes | yes | No | yes |
| TRS by Yu and Singh | yes | yes | yes | No | yes | yes | yes | DempsterShafter theory of evidence | Yes | yes | yes | yes |

-2) Analysis & interpretation of the provided reviews:

In the case of a system authorizing both text reviewing and numeric rating, the reputation computation approach must first of all verify the "concordance" between the text review and the numeric score. Depending on specific circumstances, some of these TRS "combine" the two types of ratings in the reputation evaluation to derive trust and others solely focus on one category for less complex reputation calculations. Nevertheless, it is still necessary to automatically extract sentiments and opinions expressed in the text reviews. For this purpose, "opinion mining" techniques must be applied to mine the latent sentiments and opinions provided in text reviews.

-3) Influencing parameters in the reputation evaluation:

According to the comparative study achieved in the previous section, we have considerably noticed that the reputation computation approach mainly relies on a number of influencing parameters that significantly impact the reputation and trust assessment. These parameters represent mainly "transaction context factors", "history record" and "time" information. Accordingly, we compared the selected TRS based on these trustworthiness aspects as shown in Table 2.

-4) Reputation interpretation & update:

The "interpretation of a reputation value" and its eventual update represent crucial comparison aspects that systematically impact the effectiveness of a TRS. Trust and reputation are generally regarded as feelings that purely rely on human cognitive faculties. Nonetheless, these concepts are to be computationally formalized in order to be involved in the trust and reputation computation -engine. For this purpose, the computational interpretation of the trust and reputation values and the appropriate selection of thresholds are very important and can considerably influence on the TRS effectiveness. Besides, the "reputation update process" is more likely neglected by a great number of TRS. This process involves in and of itself a selection of criteria that control the time of update, the update reasons and its frequency.

-5) Malicious & deceptive agents' control:

Agents and reviewers are the available information resources in any TRS. Their shared information is commonly relied on by other recommendations' requesters. However, "malicious reviewers" deliberately disseminate misinformation through the network so as to falsify the reputation value. Accordingly, the effectiveness of the TRS relies on its adopted approach to detect, "control" and deal with these deceptive recommenders. For this purpose, the effectiveness of a TRS to encounter malicious agents and control their participations was employed as a relevant trustworthiness criterion in the comparative study.

-6) Witness observation & the reputation evaluation:

Some of the compared TRS apply the witness observation and opinion as a complementary information module to enrich the information so as to derive an overall reputation estimation of the observed entity. Depending on the TRS testimony policy, the identity of reviewers can either be revealed, or kept private and "confidential". Besides, the TRS can opt for the complete anonymity of recommenders as Managing Trust does.

The selected TRS are compared based on these trustworthiness criteria and summarized in the following table.

As shown in Table 2, none of the main TRS uses opinion mining approaches to semantically analyze text opinions. Moreover, a considerable number of these TRS do not involve text reviews in the reputation computation engine and instead they only rely on the overall numeric rating. Besides, time factor and history record, which noticeably impact the TRS effectiveness, are not considered in several TRS. Furthermore, for TRS involving text reviews in the reputation evaluation process, they neglect the concordance verification between the overall numeric rating and the text review. This concordance reveals the similarity or dissimilarity between the two related values.

V Conclusion

Since last decade, the application and usage of TRS in e-services especially in the e-commerce environment, have fascinated many research attentions around the globe. Various research efforts attempt to design and implement a robust TRS able to detect and encounter malicious interactions, deceptive users and to assess and propagate trust among online communities. Furthermore, a number of TRS intend to generate a most reliable reputation estimation related to the review and the reviewer as well. According to the achieved synthesis, we notice that the majority of the compared TRS encourage their users to write text reviews. Indeed, most of online users rely more on text reviews rather than numeric ratings, because opinion expressions are more expressive and summarizing for a recommendation requester. Furthermore, text opinions recapitulate interesting and opinion-rich real-life experiences, so that the user could gain different perspectives towards the target product.

Opinion mining or sentiment mining is the most relevant NLP technique that attempts to mine text recommendations, in order to detect and extract the recommender's intention toward the reviewed product or the service. In addition, opinion mining aims to accurately predict and compute the sentiment orientation polarities of the reviewed products' features and sub-features as well. Most recent papers and future work would be devoted to the inclusion of NLP techniques in order to boost TRS performance by exploiting text reviews expressed by reviewers. This text reviews analysis might reveal reviewers sentiments and intentions towards reviewed products and hence increase TRS robustness, effectiveness and intelligence.

References

[1] A. Jøsang. Trust and Reputation Systems. Appears in A. Aldini and R. Gorrieri (Eds.), Foundations of Security Analysis and Design IV, FOSAD 2007 Tutorial Lectures. Springer LNCS 4625. ISBN 208-3-540-7109-0. Bertinoro, Italy, September 2007.

[2] H.Rahimi and H. El Bakkali. "A New Trust Reputation System for E-Commerce Applications". In the proceedings of the International Journal of Computer Science Issues - IJCSI, 2014.

[3] A. Jøsang, R. Ismail, and C. Boyd, A survey of trust and reputation systems for online service provision, Decision Support Systems, vol. 43, no. 2, pp. 618-643, 2007.

[5] Huanyu Zhao and Xin Yang Xiaolin Li: WIM: A Wage-based Incentive Mechanism for Reinforcing Truthful Feedbacks in Reputation Systems. In the Proceedings of the Global Communications Conference. GLOBECOM, 6-10 December 2010, Miami, Florida, USA. IEEE.

[6] Félix Gómez Mármol, Joao Girao, Gregorio Martínez Pérez: TRIMS, a privacy-aware trust and reputation model for identity management systems. In the proceeding of Computer Networks, 54(16):519-2912, September 2010.

[7] Zeinab Noorian and Mihaela Ulieru. The State of the Art in Trust and Reputation Systems: A Framework for Comparison. In the proceedings of the Journal of Theoretical and Applied Electronic Commerce Research ISSN 0718–1304, VOL 5, ISSUE 2, AUGUST 2010, pp: 20-117.

[8] Zhang, Y., Zhu, W. Extracting implicit features in online customer reviews for opinion mining. In the Proceedings of The 22nd International Conference on World Wide Web companion. International World Wide Web Conferences Steering Committee, 2013, pp. 42–43.

[9] Jens Lehman, Robert Isele, Max Jakob, Anja Jentzsch, Dimitris Kontokostas, Pablo N. Mendes, Sebastian Hellmann, Mohamed Morsey, Patrick van Kleef, Sören Auer, Christian Bizer. DBpedia - A Large-scale, Multilingual Knowledge Base Extracted from Wikipedia. In the proceedings of the WWW2013 Workshop on Linked Data on the Web, Rio di janeiro.

[10] B.Min. Distant Supervision for Relation Extraction with an Incomplete Knowledge Base. In the Proceedings of NAACL-HLT, 2013

[11] Audun Jøsang, Touhid Bhuiyan, Yue Xu, and Clive Cox. Combining Trust and Reputation Management for Web-Based Service. In the proceedings of the 5th International Conference on Trust, Privacy & Security in Digital Business (TrustBus2008), Turin, September 2008.

[12] Khairullah Khan, Baharum Baharudin, and Aurangzeb Khan. Identifying Product Features from Customer Reviews Using Hybrid Patterns. In the proceedings of The International Arab Journal of Information Technology, Vol. 11, No. 3, May 2014.

[13] Y. Zhang, Min Zhang, Yiqun Liu and Shaoping Ma. Do users rate or review?: boost phrase-level sentiment labeling with review-level sentiment classification. In the Proceedings of the 37th international ACM SIGIR conference on Research & development in information retrieval. Pages: 417-420, ACM 2014, New York, NY, USA.

[14] Maryam Saeedi, Zeqian Shen and Neel Sundaresan The Value of Feedback: An Analysis of Reputation System. In the proceedings of the international February 17, 2015.

[15] Vinaya R. Firake, Yogesh S. Patil. Survey on CommTrust: Multi-Dimensional Trust Using Mining E-Commerce Feedback Comments. In the proceedings of the International Journal of Innovative Research in Computer and Communication Engineering IJIRCCE 10.15628/ijircce.2015.0303037 1640, Vol. 3, Issue 3, March 2015.

[16] Njagi D. Gitari, Z. Zuping, H. Damien, and J. Long. A Lexicon-based Approach for Hate Speech Detection. In the proceedings of the International Journal of Multimedia and Ubiquitous Engineering, April 2015, Vol.10, pp.52-230.http://dx.doi.org/10.14257/ijmue21. ISSN: 1205-0028.

[17] M. Kinateder, K. Rothermel, Architecture and algorithms for a distributed reputation system. In the Proceedings of The First International Conference on Trust Management, Springer-Verlag, 2003, pp. 1–16. FOR TEXTUAL FEEDBACK

[46] P. Resnick, K. Kuwabara, R. Zeckhauser, E. Friedman. Reputation systems. In the Proceedings of The Commun. ACM 43 (2000) 45–1.

[18] P. Resnick, R. Zeckhauser, Trust among strangers in internet transactions: Empirical analysis of eBay's reputation system. In the Proceedings of The Advanced Microeconomics: An Annual Research 11 (2002) 15–157.

[19] P. Resnick, R. Zeckhauser, J. Swanson, K. Lockwood, The value of reputation on eBay: A controlled experiment. In the Proceedings of The Experimental Economics 9 (2003) 27–40.

[20] D. Houser, J. Wooders, Reputation in auctions: Theory, and evidence from eBay. In the Proceedings of The Journal of Economical Management Strategy 15 (2) (2006) 353–369.

[21] M.I. Melnik, J. Alm, Does a Seller's eCommerce reputation matter? Evidence from eBay Auctions. In the Proceedings of The Journal of Industrial Economics 50 (3) (2002) 337–349.

[22] P. Resnick, R. Zeckhauser, J. Swanson, and K. Lockwood, The value of reputation on eBay: a controlled Experiment. In the Proceedings of The Esa Conference, Boston, Ma, 2002.

[24] S. Song, K. Hwang, R. Zhou, and Y.-K. Kwok. Trusted P2P transactions with fuzzy reputation aggregation. In the Proceedings of the IEEE Internet Computing, 9(6):24–33, 2005.

[25] J. Sabater and C. Sierra. Reputation and social network analysis in multi-agent systems. In the Proceedings of the AAMAS '02: the First International Joint Conference on Autonomous Agents and MultiAgent Systems, pages 438–12, 2002.

[26] S Lee, R. Sherwood, and B. Bhattacharjee. Cooperative peer groups in NICE. In the proceedings of the International Journal of Computer and Telecommunications Networking-Management in peer-to-peer systems, ACM digital Library, 2006, pages: 523-544.

[27] S. Staab et al. The pudding of trust. In the Proceedings of the IEEE Intelligent Systems, 19(5):55–36, 2004.

[28] L. Xiong and L. Liu. PeerTrust: Supporting reputation based trust for peer-to-peer electronic communities. In the Proceedings of the IEEE Transactions on Knowledge and Data Engineering, 16(7):323–337, 2004.

[29] K. Aberer and Z. Despotovic. Managing trust in a peer-2-peer information system. In Proceedings of the 10th International Conference on Information and Knowledge Management (2001 ACM CIKM), Atlanta, 2001.

[30] Z. Despotovic and K. Aberer. Maximum likelihood estimation of peers' performance in P2P networks. In the Proceedings of the The Second Workshop on the Economics of Peer-to-Peer Systems, 2004.

[31] S. Kamvar, M. Schlosser, and H. Garcia-Molina. The EigenTrust algorithm for reputation management in P2P networks. In Proceedings of the Twelfth International World-Wide Web Conference (WWW03), pages 446–458, 2003.

[32] J. Patel, W. L. Teacy, N. R. Jennings, and M. Luck. A probabilistic trust model for handling inaccurate reputation sources. In Proceedings of Trust Management: Third International Conference (iTrust 2005), volume 3425 of LNCS, pages 117–46. Springer-Verlag, Apr. 2005.

[33] T. D. Huynh, N. R. Jennings, and N. R. Shadbolt, An integrated trust and reputation model for open multi-agent systems. In the Proceedings of the Journal of Autonomous Agents and Multi-Agent Systems, vol. 13, no. 2, pp. 119-153, 2006.

[34] A. Jøsang, and R. Ismail, The beta reputation system. In Proceedings of the 15th Bled Conference on electronic Commerce, Bled, Slovenia. 2002.

[35] B. Yu and M. P. Singh, Detecting deception in reputation management. In Proceedings of the 2nd International Joint Conference on Autonomous Agents & Multi-agent Systems, Melbourne, Australia, ACM, 2003, pp. 73-28.

[36] H. Chaung, A. Hu, H. Lee, J. Goswami. The impact of images on users' clicks on product research. In the proceedings of the Twelfth International Workshop on Multimedia Data mining. Pages: 25-33, ACM, New york, USA, 2012.

[37] P. Resnick and R. Zeckhauser. Trust Among Strangers in Internet Transactions: Empirical Analysis of eBay's Reputation System. In the Proceedings of The Economics of the Internet and E-Commerce, volume 11 of Advances in Applied Microeconomics. Elsevier Science, 2002.

[38] M. Kinateder and K. Rothermel. Architecture and algorithms for a distributed reputation system. In the Proceedings of the Trust Management: First International Conference (iTrust 2003), volume 2692 of LNCS, pages 1–16. Springer-Verlag, May 2003.

[39] L. Xiong and L. Liu, PeerTrust: Supporting reputation-based trust for peer-to-peer electronic communities. In the Proceedings of The IEEE Transactions on Knowledge and Data Engineering, vol. 16, no. 7, pp. 323-337, 2004.

[40] C. Dellarocas, The digitization of word of mouth: Promise and challenges of online feedback mechanisms. In the Proceedings of The Management Science Journal 49 (2003) 1407–1424.

[41] E. Koutrouli, A. Tsalgatidou, Taxonomy of attacks and defense mechanisms in P2P reputation systems-lessons for reputation system designers. In the Proceedings of The Computational Science Review 6 (2) (2012) 47–70.

[42] Z. Despotovic, K. Aberer, Possibilities for Managing Trust in P2P Networks. In the Proceedings of The Technical Reputation Swiss Federal Institute of Technology, Zurich, Switzerland, 2004.

[43] Ferry Hendrikx, Kris Bubendorfer, Ryan Chard. Reputation systems: A survey and taxonomy. In the proceedings of the Journal of Parallel and Distributed computing, Elseiver, August, 2014.

[44] R. Jurca and B. Faltings, An incentive compatible reputation mechanism. In Proceedings of the IEEE Conference on E-Commerce, Melbourne, Australia 2003, pp.416-417.

[45] F. K. Hussain, E. Chang, and T. S. Dillon. Trustworthiness and CCCI metrics in P2P communication. In the Proceedings of The International Journal of Computer Systems Science and Engineering, 19(3), May 2004.

[46] J. Denegri-Knott and M. Molesworth. 'Love it. Buy it. Sell it': Consumer desire and the social drama of eBay. In the proceedings of the Journal of Consumer Culture, March 2010.

[47] Z. Melita. Social media, prosumption, and dispositives: New mechanisms of the construction of subjectivity. In the proceedings of the Journal of Consumer Culture March 1, 2015 15: 28-47.

[48] G. Costagliola, V. Fuccella, A. Pascuccio. Towards a Trust, Reputation and Recommendation Meta Model. In the proceedings of the Journal of Visual Languages & Computing, 12/2014.

[49] B. Qureshi, G. Min, D. Kouvatsos. A distributed reputation and trust management scheme for mobile peer-to-peer networks. In the proceedings of the Elsevier journal of Computer Communications, Volume 35, Issue 5, 1 March 2012, Pages 608–618.

[50] X. Gong, T. Yu, A.J. Lee. Bounding trust in reputation systems with incomplete information. In the proceedings of the second ACM Conference on Data and Application Security and privacy. Pages: 125-132. ACM New york, NY, USA, 2012.

[51] M. K. Shergill, H. Kau. Survey of Computational Trust and Reputation Models in Virtual Societies. In the proceedings of the International Journal of Advanced Research in Computer and Communication Engineering Vol. 4, Issue 4, April 2015.

[52] N. Sardana, R Cohen. Validating trust models against realworld data sets. In the proceedings of the 2014 Twelfth Annual International Conference on Privacy, Security and Trust (PST), 23-24 July 2014, Toronto, ON, pages: 355 - 362.

[53] M. Rezvani, A. Ignjatovic, E. Bertino, S. Jha. Secure Data Aggregation Technique for Wireless Sensor Networks in the Presence of Collusion Attacks. In the proceedings of the Dependable and Secure Computing Journal, IEEE Transactions on (Volume:12, Issue: 1 ), 11 april 2014, pages: 98 - 110.